%% file: main.tex
\documentclass[11pt,a4paper]{article}
\usepackage{stysty}
\usepackage{amsmath}
\usepackage[height=8.85in,width=6.45in]{geometry}
\usepackage{cite}

\title{Matrix product state classification of 1D multipole symmetry protected
topological phases}
\author[1]{Takuma Saito}
\author[2,3]{Weiguang Cao}
\author[4]{Bo Han}
\author[5,1]{Hiromi Ebisu}
\affil[1]{Yukawa Institute for Theoretical Physics, Kyoto University, Kyoto 606-8502, Japan}
\affil[2]{Center for Quantum Mathematics at IMADA, Southern Denmark University, Campusvej 55, 5230 Odense, Denmark}
\affil[3]{Niels Bohr International Academy, Niels Bohr Institute, University of Copenhagen, Blegdamsvej 17DK-2100 Copenhagen, Denmark}
\affil[4]{Institute for Theoretical Physics, University of Cologne, Physik, Zülpicher Str. 77, 50937 Köln, Germany}
\affil[5]{Interdisciplinary Theoretical and Mathematical Sciences Program (iTHEMS)~RIKEN, Wako 351-0198, Japan}

\newcommand{\ueq}{\overset{\cdot}{=}}

\theoremstyle{plain}
\newtheorem{propo}{Proposition}

\begin{document}

\maketitle

\begin{abstract}
Spatially modulated symmetries are one of the new types of symmetries whose symmetry actions are position dependent. Yet exotic phases resulting from these spatially modulated symmetries are not fully understood and classified. In this work, we systematically classify one dimensional bosonic symmetry protected topological phases protected respecting multipole symmetries by employing matrix product state formalism. The symmetry action induces projective representations at the ends of an open chain, which we identify via group cohomology. In particular, for 
$r$-pole symmetries, for instance,  $r$ = 0 (global), 1 (dipole), and 2 (quadrupole), the classification is determined by distinct components of second cohomology groups that encode the boundary projective representations.
\end{abstract}

\tableofcontents

\section{Introduction}
Classification of phases of matter is one of the main themes in condensed matter physics. In this viewpoint, topological phases of matter are of particular importance as they are beyond the standard paradigm of the Ginzburg-Landau formalism. 
 New types of phases of matter include \textit{symmetry protected topological~(SPT) phases}, which are invertible phases with unique gapped state protected by global symmetries~\cite{GuWen0903,Senthil1405,PollmannBergTurnerOshikawa0909,ChenLiuWen1106,LevinGu1202,YaoRyu1202,ChenLuVishwanath1303}.
At the first glance, these phases  look trivial in the bulk as they are merely gapped phases with unique ground state. Yet, there are rich boundary physics, such as nontrivial boundary states forming projective representations of a global symmetry and chiral edge modes. In particular, the discovery of topological insulators has spurred new classification schemes by employing group cohomology~\cite{ChenGuWen1008,SchuchPerezGarciaCirac1010,ChenGuLiuWen1106,ElseNayak1409,Ogata2110}. 
Furthermore, SPT phases play pivotal roles in understanding 't Hooft anomaly in view of high energy physics~\cite{Wen2013sptanomaly}. 

\par
On a different route, there has been growing interest in extending the concept of symmetry,  which is the most fundamental guiding principle of physics. One example is the \textit{modulated symmetry}, whose actions on local degrees of freedom are position-dependent. These spatially modulated symmetries were originally proposed in fracton phases, where there are excitations with mobility constraints~\cite{Chamon0404,BravyiLeemhuisTerhal1006,Haah1101,Yoshida1302,VijayHaahFu1505,VijayHaahFu1603,Pretko1604,Pretko1606,PremHaahNandkishore1702,VijayFu1706,SlagleKim1708,PremPretkoNandkishore1709,SlagleKim1712}.
(For reviews on fractons, see e.g.,
\cite{NandkishoreHermele1803Fractons,PretkoChenYou2001}). However, the systematic classification of SPT phases protected by spatially modulated symmetries has not been fully understood. 

In the present work, we initialize the systematic classification by focusing on the simplest example of the modulated symmetries, i.e. \textit{multipole symmetries}, associated with conservation of multipoles, such as dipole symmetries~\cite{Seiberg:2019vrp,griffin2015scalar,Pretko:2018jbi,PhysRevX.9.031035,Gorantla:2022eem}.
The multipole symmetries find various applications as they impose substantial constraints on dynamics of a system, leading to intriguing physical consequences.
For instance, there are rich phases in a new type of Bose-Hubbard model with multipole symmetries~\cite{LakeHermeleSenthil2201,LakeLeeHanSenthil2210,ZechmannAltmanKnapFeldmeier2210}. 
Also, systems with multipole symmetries lead to the Hilbert space fragmentation and breaking of ergodicity
\cite{SalaRakovszkyVerresenKnapPollmann1904,MoudgalyaPremNandkishoreRegnaultBernevig1910,KhemaniHermeleNandkishore1910}.
There are considerable amount of interest in establishing topologically ordered phases with the multipole symmetries~\cite{PhysRevB.106.045145,2023foliated}, especially in their connections to symmetry enriched topological (SET) phases~\cite{2024multipole,PaceLamAksoy2409},  the Lieb-Schultz-Mattis constraints~\cite{AksoyMudryFurusakiTiwari2308LSManomalies&webdualities,PaceLamAksoy2409,2025lsm}, and exotic ground state degeneracy depending on the system size, which is a manifestation of the UV/IR mixing~\cite{Pace:2022wgl}.

\par
In accordance with the development of new types of symmetries, establishing new classification schemes of phases with these symmetries has attracted more attention and become an active area of research. For example, in recent years, classification of SPT phases with other generalized symmetries, like subsystem symmetry~\cite{DevakulWilliamsonYou1808,Devakul1812,DevakulShirleyWang1910,JiaJia2505}, fractal symmetry~\cite{Devakul:2018lmi}, and categorical symmetry (a.k.a. noninvertible symmetry)\cite{ThorngrenWang1912,KongLanWenZhangZheng2003,ChatterjeeWen2205,MoradiMoosavianTiwari2310,BhardwajPajerSchaferNamekiTiwariWarmanWu2408,BhardwajSchaferNamekiTiwariWarman2502,Cao:2025qhg,ParayilMana:2025nxw,Furukawa:2025flp,doi:10.1142/S0217751X25480021}, has been intensively studied.
This work aims to establish a classification of SPT phases with multipole symmetries. 
This problem was partly addressed by an explicit lattice model construction of the $(1+1)$d SPT phases with dipole symmetry~\cite{Lam2311} and introducing more generic SPT phases in the language of foliated gauge field theories in view of anomaly inflow for modulated symmetries~\cite{anomaly_2024}~\footnote{See e.g.,~\cite{Burnell:2021reh,Yamaguchi:2021xeq,Honda:2022shd,Ding:2024jxe} and~\cite{Casasola:2024esm} for discussion on anomaly inflow for subsystem symmetries and fractal symmetries, respectively.}.
Further, via defect network constructions, 
classification of the SPT phases with dipole symmetries in $(2+1)$d has been considered~\cite{Bulmash2508,Yao25toappear}.
However, exploration of the SPT phases with multipole symmetries, especially the lattice model construction of them, has yet to be fully made.\par
In this paper, by utilizing matrix product state~(MPS) formalism\cite{Fannes:1990ur,SchuchPerezGarciaCirac1010,Lam2311}, we demonstrate an explicit lattice model construction of SPT phases with Abelian multipole symmetries in $(1+1)$d and systematically classify distinct phases. 
Specifically, by 
constructing MPS with Abelian $r$-pole ($r\in\mathbb{N}$) symmetries and investigating commutation relations of symmetry actions on the boundaries, we find that distinct SPT phases are labeled by the ratio between cohomology groups, as summarized in Eq.~\eqref{eq: classification of degree-r multipole SPT phases} and equivalently in  Eq.~\eqref{eq: classification of degree r multipole in fractional way}.
Our results would hold values for 
classifications of topological phases with multipole symmetries and also for understanding topological phases protected by other types of modulated symmetries.
\par
The rest of this work is organized as follows:
In Sec.~\ref{sec2}, we present MPS formalism to build up SPT phases with abelian multipole symmetries. In Sec.~\ref{sec3}, based on the model that we construct, we give classification of the SPT phases. Finally, in Sec.~\ref{sec4}, we conclude our work with a few remarks. Technical issues are relegated to appendices.

\section{MPS formalism of multipole SPT phases}\label{sec2}
In this section, following~\cite{Lam2311}, we introduce MPS formalism of the multipole SPT phases, and examine the multipole symmetry actions on the boundaries, which are crucial to deriving the classification of multipole SPT phases. 
\subsection{Symmetries in translation and multipole symmetric systems}
To start, 
we consider a finite abelian group $G$.
Let $\Lambda$ be a spin chain, $\bbC^{|G|}$ be a Hilbert space on each site, and for all $g\in G$, let $g_x$ be an on-site unitary operator acting on the site $x\in\Lambda$.
We define rank-$r$ multipole symmetry in the following way: 
\begin{equation}
    U_g^{(r)}
    :=
    \prod_{x\in\Lambda}
    (g_x)^{\frac{1}{r!}x(x+1)\cdots(x+r-1)}, \quad r\in \mathbb N
    .
    \label{eq: multipole symmetry definition}
\end{equation}
Note that the exponent $x(x+1)\cdots(x+r-1)/r!=\binom{x+r-1}{r}$ is an integer, and that any integer-valued polynomial with rank $r_0$ is a superposition of $\{\binom{x+r-1}{r}\}_{r=0}^{r_0}$ with integer coefficients.
The key feature of the global $r$-pole symmetry~\eqref{eq: multipole symmetry definition} is that lower rank modulated symmetries are generated by acting a translation operator $T$.  To be more explicit, when the system has rank-$r$ multipole symmetry and translation symmetry $T$, it also has lower-rank multipole symmetries. For example
\begin{equation}\label{eq: multipole symmetry under translation_1}
   U_g^{(r)} T\left(U_g^{(r)}\right)^{-1}T^{-1} =
    \prod_{x\in\Lambda}
    g_x^{\binom{x+r-1}{r}-\binom{x+r-2}{r}}
    =
    \prod_{x\in\Lambda}
    g_x^{\binom{x+r-2}{r-1}}
    =
    U_g^{(r-1)}
    .
\end{equation}
Inductively, we can show that the system with rank-$r$ multipole symmetry and translation symmetry $T$ has all lower 
ranks of multipole symmetries, including uniform symmetry $U_g^{(0)}:=\prod_{x\in\Lambda} g_x$.

\subsection{Symmetry actions in the bulk}


Let an injective MPS
\begin{equation}
    \label{eq: MPS}
    \ket{\psi}
    =\sum_{\textbf{h}}\text{Tr}[\cdots A^{h_{x}}A^{h_{x+1}}\cdots]\ket{\textbf{h}}:=
    \cdots
    \vcenter{\hbox{

\input{img/MPS.tex}
    }}
    \cdots
\end{equation}
be a SPT ground state of a system with rank-$r$ multipole symmetry $U_g^{(r)}$ and translation symmetry $T$ in the periodic boundary condition\footnote{
Throughout this paper, we focus on the case where $G$ is Abelian, $G=\prod_{i=1}^k\mathbb{Z}_{N_i}$. 
To keep the multipole symmetry for periodic boundary condition, we need to require that the lattice size $L$ is divisible by ${N_i}$~$(i=1,\cdots,k)$}.
The MPS $\ket{\psi}$ is constructed by tracing over the internal indices $p,q$ of MPS tensors $A_{p,q}^{h}$, and the symmetry operators can act on the physical index $h$. In the diagram representation, we suppress the indices and simply use $A^{(x)}$ when we would like to emphasize the locations of the MPS tensors.
As we mentioned above, the system also has the rank-0 symmetry $\{U_g^{(0)}|g\in G\}$.
The fundamental theorem of MPS \cite{Perez-GarciaVerstraeteWolfCirac0608,PerezGarciaWolfSanzVerstraeteCirac0802} guarantees that $g_x$ acts on an MPS tensor $A$ as
\begin{equation}
    \vcenter{\hbox{

\input{img/monopole-MPS.tex}
    }}
    \ueq
    \vcenter{\hbox{
        \input{img/monopole-act-result.tex}
    }}
    =:
    \qty(X_g^{(0)})^\dag
    A
    X_g^{(0)}
\end{equation}
where $X_g^{(0)}$ is a unitary operator on virtual spaces.
Here $\ueq$ means that the left-hand side is equal to the right-hand side up to U(1) phase factors.
To generalize the argument to the multipole symmetry, one can think of how the dipole symmetry acts on the MPS. Indeed, 
\begin{equation}
    \begin{split}
        &
        \cdots
        \vcenter{\hbox{
            \input{img/dipole-MPS.tex}
        }}
        \cdots
        \\
        &\qquad
        \ueq
        \cdots
        \qty(X_g^{(0)})^{-x}
        A^{(x)}
        \qty(X_g^{(0)})^{x}
        \qty(X_g^{(0)})^{-(x+1)}
        A^{(x+1)}
        \qty(X_g^{(0)})^{(x+1)}
        \cdots
        \\
        &\qquad
        =
        \cdots
        \qty(X_g^{(0)})^{-1}
        A^{(x)}
        \qty(X_g^{(0)})^{-1}
        A^{(x+1)}
        \cdots
        .
    \end{split}
\end{equation}
Since we assumed $\ket{\psi}$ is a unique ground state, the right hand side is equivalent to $\ket{\psi}$ up to a~U(1) phase factor.
There is a unique $X_g^{(1)}$ such that
\begin{eqnarray}
    \vcenter{\hbox{
        \input{img/mono-gauge.tex}
    }}
    \ueq
    \vcenter{\hbox{
        \input{img/dipole-newgauge.tex}
    }}
    .\label{dipole}
\end{eqnarray}

\begin{figure}
    \centering
\includegraphics[width=1.0\linewidth]{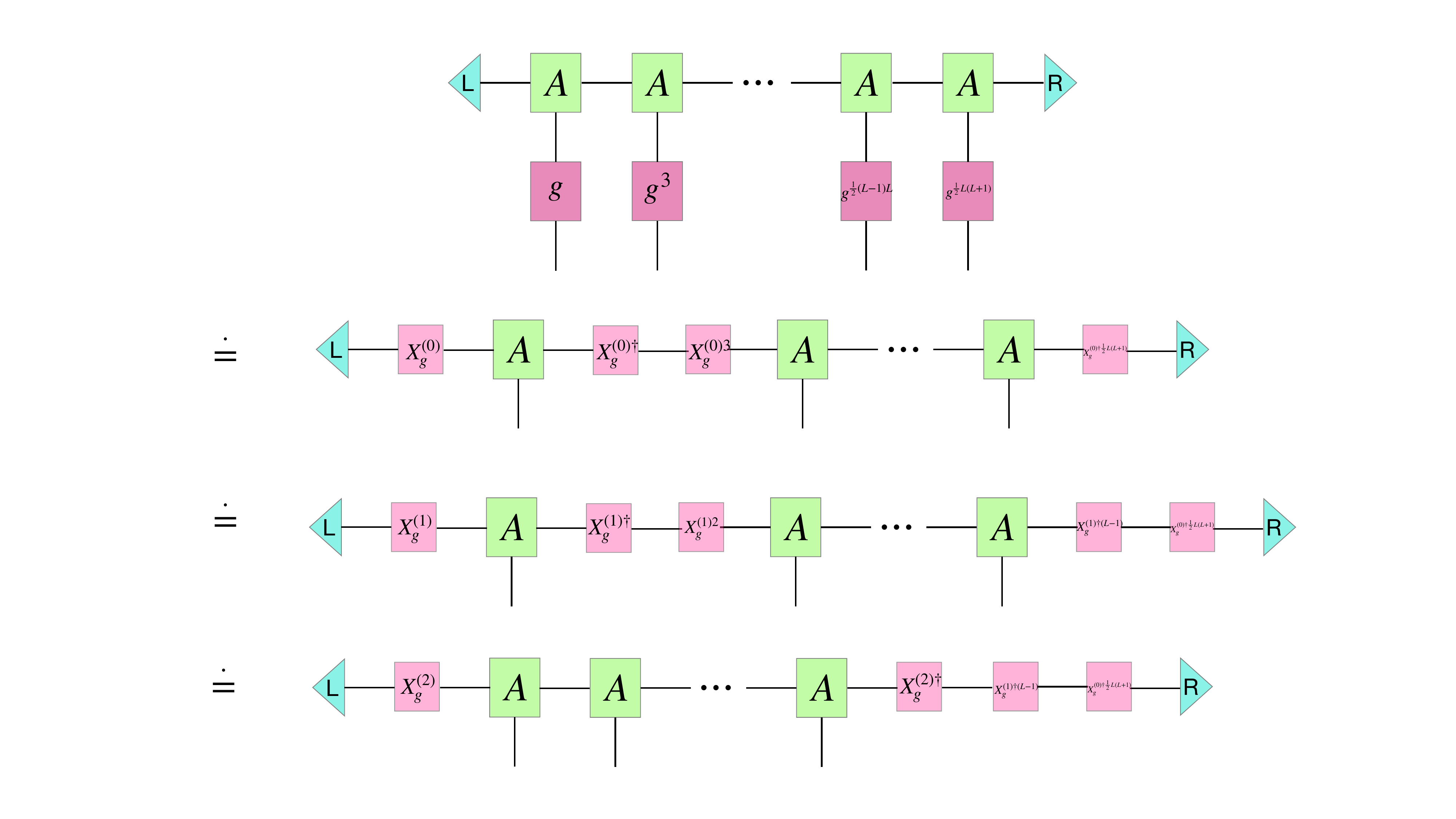}
    \caption{The action of quadrupole symmetry~$\text{Qu} = \prod_j g^{\frac{1}{2}j(j+1)}$ on the MPS. 
    %
    }
    \label{fig:placeholder}
\end{figure}

Once we have~\eqref{dipole}, the action of multipole symmetries with rank $r~(r\geq 2)$ on the MPS can be derived
inductively. 
Indeed, one can show that 
$U^{(k+1)}$ gives level-$(k+1)$ unitary $X_g^{(k+1)}$.
\begin{equation}
    \label{mps k}
    \vcenter{\hbox{
        \input{img/k-gauge.tex}
    }}
    \ueq
    \vcenter{\hbox{
        \input{img/k-new-gauge.tex}
    }}.\label{mps k}
\end{equation}
In \cref{sec: proof of k gauge into k+1 gauge}, we review the rigorous proof for~\eqref{mps k} given in \cite{BridgemanChubb1603,Lam2311}. See also Fig.~\ref{fig:placeholder} for an example of the action of the quadrupole symmetry on the MPS.

\section{Classification of rank-$r$ multipole SPT phases}\label{sec3}
In this section, we derive the classification of the multipole SPT phases. Based on the construction presented in the previous section, we investigate how multipole symmetry operator~$U_g^{(k)}$ ($0\leq k\leq r$, $g\in G$) acts on the MPS with a boundary. In particular, we examine the projective realization of multipole symmetries~$U_g^{(k)}$ and $U_h^{(l)}$ ($0\leq k,l\leq r$, $g,h\in G$) on the boundary, from which we classify the multipole SPT phases. 

\subsection{Symmetry actions on the boundary}

Nontrivial edge modes are one of the characteristics of SPT phases.
In this paper, we classify SPT phases by investigating edge modes in open boundary conditions.
Iterative use of~\eqref{mps k} leads to the action of symmetry operator $U_g^{(k)}$  on an open spin chain with system size $L$ as
\begin{equation}
    \label{eq: boundary action}
    U_g^{(k)}\ket{\psi}
    \ueq
    \qty(X_g^{(k)})^\dag
    A^{(1)}A^{(2)}\cdots A^{(L)}
    \qty(
        \prod_{n=0}^{k}
        \qty(X_g^{(n)})^{(L+k-n-1)!/(k-n)!(L-1)!}
    )
    .
\end{equation}
Using this relation, we can investigate boundary of the MPS to classify the $r$-pole SPT phases.  
Since unitaries  $X_g^{(k)}$ ($0\leq k\leq r$, $g\in G$) form $(r+1)$-flavor projective representation of $G$, naively they are expected to be classified by the second group cohomology $H^2(G^{\times(r+1)},U(1))$.
However, 
we have additional constraints on $X_g^{(k)}$ coming from the commutation relations of the projective representations on the boundaries, necessitating taking quotient of some of cohomology groups. 
Because $U_g^{(k)}$ and $U_g^{(l)}$ commute as operators acting on the whole chain, the phase factors from commutation of unitaries on each boundary should compensate.

Let the commutation relation of unitaries be
\begin{equation}
    \label{eq: commutation relation of single gauge}
    X_g^{(k)}X_h^{(l)}
    =
  e^{i\theta_{g,h}(k,l)}
    X_h^{(l)}X_g^{(k)}
    . 
\end{equation}
The phase factor is gauge invariant, i.e., it is invariant under the transformation $X_g^{(k)}\mapsto e^{i\varphi_g^{(k)}}X_g^{(k)}$.
By definition, we have
\begin{equation}
    \label{eq: ghkl is -hglk}
    e^{i\theta_{g,h}{(k,l)}}
    =
    e^{-i\theta_{h,g}{(l,k)}}.
\end{equation}
It is known that $e^{i\theta_{g,h}(k,l)}$ is a bicharacter of $G$, and the set of bicharacters $\{e^{i\theta_{g,h}(k,l)}\mid 0\leq k,l\leq r,\;g,h\in G\}$ is isomorphic to the second group cohomology $H^2(G^{\times(r+1)},U(1))$ (see e.g. \cite{Tambara0012}).
To proceed, 
let $[U,V]$ be the commutator of two unitary operators $U$ and~$V$, viz,~\footnote{Throughout this paper, we define the commutator in this way, in contrast to the common commutator $[A,B] = AB - BA$. Our convention makes the topological invariant manifest. }
\begin{equation}
    [U,V]:=UVU^{-1}V^{-1}.
\end{equation}
We investigate the U(1) phase on both ends of the system. 
On the left edge, we have
\begin{equation}
    \qty[
        \qty(X_h^{(l)})^\dag,
        \qty(X_g^{(k)})^\dag
    ]
    =
    \qty[
        \qty(X_g^{(k)}),
        \qty(X_h^{(l)})
    ]^\dag
    =
    e^{-i\theta_{g,h}(k,l)}
\end{equation}
whereas on the right edge, we obtain
\begin{equation}
    \begin{split}
        &
        \qty[
            \prod_{m=0}^{k}
            \qty(X_g^{(m)})^{(L+k-m-1)!/(k-m)!(L-1)!},
            \prod_{n=0}^{l}
            \qty(X_h^{(n)})^{(L+l-n-1)!/(l-n)!(L-1)!}
        ]
        \\
        &\qquad
        =
        \prod_{m=0}^{k}
        \prod_{n=0}^{l}
        \exp(
            i\theta_{g,h}(m,n)
            \frac{(L+k-m-1)!}{(k-m)!(L-1)!}
            \frac{(L+l-n-1)!}{(l-n)!(L-1)!}
        )
    \end{split}
\end{equation}
The whole phase factor satisfies
\begin{equation}
    \label{eq: commutativity condition in rough form}
    1
    =
    \exp(
        \sum_{m=0}^{k}
        \sum_{n=0}^{l}
        i\theta_{g,h}(m,n)
        \binom{L+k-m-1}{k-m}
        \binom{L+l-n-1}{l-n}
        -
        i\theta_{g,h}(k,l)
    )
\end{equation}
for all $g,h\in G$, $0\leq k,l\leq r$ and for all system size $L$.


The identity~\eqref{eq: commutativity condition in rough form} can be simplified using the following proposition:
\begin{propo}
    \label{prop:simplified commutativity condition}
    Let the system has rank-$r$ multipole symmetry $U_g^{(r)}$ and translation symmetry $T$.
    The following two conditions are equivalent:
    \begin{enumerate}
        \item \label{item: commutativity condition in rough form}
            The commutativity condition \eqref{eq: commutativity condition in rough form} is satisfied for all $g,h\in G$, $0\leq k,l\leq r$ and for all system size $L$.
        \item \label{item: commutativity condition in simplified form}
            For all $g,h\in G$, 
            \begin{equation}
                \label{eq: commutativity condition in simplified form}
                \begin{cases}
                    e^{i\theta_{g,h}(k,l)}=1 & (k+l<r)
                    \\
                    e^{i\theta_{g,h}(k,l)}=e^{i\theta_{g,h}(k+1,l)}e^{i\theta_{g,h}(k,l+1)} & (k,l<r)
                \end{cases}
            \end{equation}
    \end{enumerate}
\end{propo}
We prove this proposition in \cref{sec: proof of prop. simplified commutativity condition}. 
In summary, we have constructed a MPS which respects $r$-pole symmetry. By introducing a boundary, we obtain relations between U(1) phases associated with symmetry operators~\eqref{eq: commutativity condition in simplified form}. In the following, we make use of these relations to classify the SPT phases protected by multipole symmetry.

\subsection{The case with $r=3$}
For illustrative purposes, we first focus on the case with $r=3$. In this case, we summarize U(1) phases in~\eqref{eq: commutativity condition in simplified form} in Table~\ref{tab:r3 thetas}.
\begin{table}
    \centering
    \caption{U(1) phases~$\theta_{g,h}(k,l)$ in the case of $r=3$, which satisfy~\eqref{eq: commutativity condition in simplified form}. The entries of the right-most column determine other values in the tables. To wit, there are four independent phases marked by red color which generate other phases. 
    }
    \label{tab:r3 thetas}
    \begin{tabular}{c|cccc}
    \hline
        $k\backslash l$ & 0 & 1 & 2 & 3
        \\
        \hline
        0 & 0 & 0 & 0 & $\red{\theta_{g,h}(0,3)}$
        \\
        1 & 0 & 0 & $-\theta_{g,h}(0,3)$ & $\red{\theta_{g,h}(1,3)}$
        \\
        2 & 0 & $\theta_{g,h}(0,3)$ & $-\theta_{g,h}(0,3)-\theta_{g,h}(1,3)$ & $\red{\theta_{g,h}(2,3)}$
        \\
        3 & $-\theta_{g,h}(0,3)$ & $2\theta_{g,h}(0,3)+\theta_{g,h}(1,3)$ & $-\theta_{g,h}(0,3)-\theta_{g,h}(1,3)-\theta_{g,h}(2,3)$ & $\red{\theta_{g,h}(3,3)}$\\
        \hline
    \end{tabular}
\end{table}
A key insight of these phases is that most of them are not independent, that is, once we fix the value of a phase, the other phases are determined by the conditions of~\eqref{eq: commutativity condition in simplified form}.
Indeed, 
up to the minus sign, 
the entry $\theta_{g,h}(3,0)$, $\theta_{g,h}(2,1)$, $\theta_{g,h}(1,2)$ 
are identical to $\theta_{g,h}(0,3)$ due to the condition~\eqref{eq: commutativity condition in simplified form} jointly with the fact that 
$\theta_{g,h}(0,3)=-\theta_{g,h}(3,0)=\theta_{h,g}(0,3)$ which is derived from \eqref{eq: ghkl is -hglk}. 
Likewise, from the condition~\eqref{eq: commutativity condition in simplified form},
we also check whether the four independent phases are symmetric or antisymmetric under exchanging group elements, $g$ and $h$. 
From~\eqref{eq: commutativity condition in simplified form}, jointly with~\eqref{eq: ghkl is -hglk}, 
one finds that 
\begin{eqnarray}
    \theta_{g,h}(0,3)-  \theta_{h,g}(0,3)=0,
\end{eqnarray}
implying antisymmetric part of $\theta_{g,h}(0,3)$ is trivial. Further, it can be verified that 
\begin{eqnarray}
   \theta_{g,h}(1,3)+\theta_{h,g}(1,3)=-2\theta_{g,h}(0,3).
\end{eqnarray}
This indicates that once we give a value to $\theta_{g,h}(0,3)$, the value of 
symmetric part of $\theta_{g,h}(1,3)$ is identified. 
In the same manner, 
from
\begin{eqnarray}
    \theta_{g,h}(2,3)-\theta_{h,g}(2,3)=-\theta_{g,h}(0,3)-\theta_{g,h}(1,3),
\end{eqnarray}
it follows that once we give values to $\theta_{g,h}(0,3)$ and $\theta_{g,h}(1,3)$, the value of symmetric part of $  \theta_{g,h}(2,3)$ is determined. 
From~\eqref{eq: ghkl is -hglk}, one has $\theta_{g,h}(3,3)+\theta_{h,g}(3,3)=0$, which implies that symmetric part of $\theta_{g,h}(3,3)$ becomes trivial.

To see how to label
symmetric/antisymmetric part
of a phase, 
we consider the commutation of representation $X_g^{(k)}X_g^{(3)}$ for $k=0,1,2,3$, which corresponds to diagonal subgroup $\text{diag} (G\times G)$:
\begin{equation}
    \begin{split}
        \qty[
            X_g^{(k)}X_g^{(3)},
            X_h^{(k)}X_h^{(3)}
        ]
        &=
        e^{i\theta_{g,h}(k,k)}
        e^{i\theta_{g,h}(3,3)}
        e^{i\theta_{g,h}(k,3)+i\theta_{g,h}(3,k)}
        \\
        &=
        e^{i\theta_{g,h}(k,k)}
        e^{i\theta_{g,h}(3,3)}
        \prod_{k'+l'<k+3}e^{in_{k',l'}\theta_{g,h}(k',l')}
        ,
    \end{split}
\end{equation}
with some integers $\{n_{k',l'}\}$.
Since commutation of representation on diagonal subgroup is represented only by lower ranks, we have to quotient out the classification by $H^2(G,U(1))$ corresponding to representation of diagonal symmetry $G$.

To recap the argument, the multipole SPT phases with $r=3$, are described by four distinct phases, 
symmetric part of $\theta_{g,h}(0,3)$ and $\theta_{g,h}(2,3)$, and antisymmetric part of $\theta_{g,h}(1,3)$ and $\theta_{g,h}(3,3)$. Generally, 
a phase $\theta_{g,h}(k,l)$ is characterized by  $H^1(G,H^1(G,U(1)))~(k\neq l)$ and~$H^2(G,U(1))~(k=l)$. Therefore, symmetric part of $\theta_{g,h}(k,l)~(k\neq l)$ is labeled by
\begin{eqnarray}
        \frac{H^1(G,H^1(G,U(1)))}{H^2(G,U(1))},
\end{eqnarray}
where we have mod out the diagonal subgroup. Also, antisymmetric part of $\theta_{g,h}(k,l)~(k\neq l)$ is described by
\begin{eqnarray}
       \frac{H^1(G,H^1(G,U(1)))}{{H^1(G,H^1(G,U(1)))}/{H^2(G,U(1))}}
    \cong 
    H^2(G,U(1))
\end{eqnarray}
whereas $\theta_{g,h}(k,k)$ is by $H^2(G,U(1))$.
Overall, the multipole SPT phases with $r=3$ are labeled by
\begin{eqnarray}
      \qty[
                H^2(G,U(1))
                \oplus
                \frac{H^1(G,H^1(G,U(1)))}{H^2(G,U(1))}
            ]^{\oplus 2}.
\end{eqnarray}
\subsection{Generic value of $r$}
We turn to classification of the multipole SPT phases with generic value of $r$. Since discussion closely parallels the one in the previous subsection, we present it succinctly.  
Iterative use of~\eqref{eq: commutativity condition in simplified form} and~\eqref{eq: ghkl is -hglk} leads to $(r+1)$~independent phases $\theta_{g,h}(a,r)~(0\leq a\leq r)$
with other phases either determined by linear combination of them or trivial. Furthermore, 
symmetric part of $\theta_{g,h}(a,r)~(a=r,r-2,\cdots,)$ with respect to exchanging group elements, $g$ and $h$
are trivial whereas antisymmetric part of $\theta_{g,h}(a,r)~(a=r-1,r-3,\cdots,)$ become trivial. 
\par
In the case of $k+l<2r$, $\theta_{g,h}(k,l)$ is symmetric if $k+l$ is odd under exchange of $g$ and $h$, and antisymmetric if $k+l$ is even.
Further, from~\eqref{eq: ghkl is -hglk}, $\theta_{g,h}(r,r)$ is antisymmetric.
These conditions give explicit classification of multipole SPTs.
To see this, we resort to the Künneth formula
\begin{equation}
    H^2(G^{\times(r+1)},U(1))
    =
    [H^2(G,U(1))]^{\oplus(r+1)}
    \oplus
    [H^1(G,H^1(G,U(1)))]^{\oplus r(r+1)/2}.
\end{equation}
This implies that equivalence class $[\theta_{g,h}(k,l)]$ is in $H^2(G,U(1))$ if $k=l$, and in $H^1(G,H^1(G,U(1)))$ if $k\neq l$.
Above consideration indicates that symmetric part of $\theta_{g,h}(k,r)$ is trivial up to $\theta(k',l')$ with $k'+l'<k+r$ if $k+r<2r$ is even.
Here we consider the commutation of representation $X_g^{(k)}X_g^{(r)}$ which corresponds to diagonal subgroup $\text{diag} (G\times G)$:
\begin{equation}
    \label{eq: commutation of X_g(0)X_g(r) in case r is odd}
    \begin{split}
        \qty[
            X_g^{(k)}X_g^{(r)},
            X_h^{(k)}X_h^{(r)}
        ]
        &=
        e^{i\theta_{g,h}(k,k)}
        e^{i\theta_{g,h}(r,r)}
        e^{i\theta_{g,h}(k,r)+i\theta_{g,h}(r,k)}
        \\
        &=
        e^{i\theta_{g,h}(k,k)}
        e^{i\theta_{g,h}(r,r)}
        \prod_{k'+l'<k+r}e^{in_{k',l'}\theta_{g,h}(k',l')}
        , 
    \end{split}
\end{equation}
with some integers $\{n_{k',l'}\}$.
Since commutation of representation on diagonal subgroup is represented only by lower ranks and $\theta_{g,h}(r,r)$, we have to quotient out the classification by $H^2(G,U(1))$ corresponding to representation of diagonal symmetry $G$.
Hence, symmetric part is given by
\begin{equation}
    \frac{H^1(G,H^1(G,U(1)))}{H^2(G,U(1))},
\end{equation}
whereas, the antisymmetric part is 
\begin{equation}
    \frac{H^1(G,H^1(G,U(1)))}{{H^1(G,H^1(G,U(1)))}/{H^2(G,U(1))}}
    \cong 
    H^2(G,U(1)).
    \end{equation}
The phase $\theta(r,r)$ is classified by $H^2(G,U(1))$.

In summary, classification of rank-$r$ multipole SPT phases is written as the follows:
\begin{equation}
    \label{eq: classification of degree-r multipole SPT phases}
    \boxed{
    \begin{split}
        \mathcal{C}^{(r)}(G)
        &\cong
        H^2(G,U(1))
        \oplus
        \frac{H^1(G,H^1(G,U(1)))}{H^2(G,U(1))}
        \oplus
        H^2(G,U(1))
        \oplus
        \cdots
        \\
        &\cong
        \begin{cases}
            \qty[
                H^2(G,U(1))
                \oplus
                \frac{H^1(G,H^1(G,U(1)))}{H^2(G,U(1))}
            ]^{\oplus m} 
            & (r=2m-1)
            \\
            [H^2(G,U(1))]^{\oplus (m+1)}
            \oplus 
            \qty[\frac{H^1(G,H^1(G,U(1)))}{H^2(G,U(1))}]^{\oplus m} 
            & (r=2m)
        \end{cases}
    \end{split}}
\end{equation}
The result~\eqref{eq: classification of degree-r multipole SPT phases} can be rewritten as the follows via Künneth formula:
\begin{equation}
    \label{eq: classification of degree r multipole in fractional way}
    \boxed{
    \mathcal{C}^{(r)}(G)
    =
    \frac{H^2(G^{\times(r+1)},U(1))}{[\mathcal{C}^{(0)}(G)]^{\oplus2}[\mathcal{C}^{(1)}(G)]^{\oplus2}\cdots[\mathcal{C}^{(r-1)}(G)]^{\oplus2}}}
\end{equation}
The formula~\eqref{eq: classification of degree r multipole in fractional way} implies that multipole SPT phases with $r=r_0$ can be classified by induction from the ones with $r=0,1,2,\cdots,r_0-1$.  
Note that in the case of $r=0$, the result becomes the usual classification of SPT phases with $G$ symmetry, and the case with $r=1$ agrees with the classification as in \cite{Lam2311}. Our result is compatible with a defect network construction studied in~\cite{Bulmash2508,Yao25toappear} which focuses only on $G=\mathbb{Z}_N$.
By the fundamental theorem of finite abelian group, $G$ is represented in the form of $G=\prod_{i\in I}\mathbb{Z}_{N_i}$.
In this case,
\begin{equation}
    \begin{split}
        &
        H^2(G,U(1))
        \cong
        \prod_{i<j}\mathbb{Z}_{\gcd(N_i,N_j)}
        ,
        \\
        &
        H^1(G,H^1(G,U(1)))
        \cong
        \prod_{i,j}\mathbb{Z}_{\gcd(N_i,N_j)}
        ,
        \\
        &
        \frac{H^1(G,H^1(G,U(1)))}{H^2(G,U(1))}
        \cong
        \prod_{i\leq j}\mathbb{Z}_{\gcd(N_i,N_j)}
        .
    \end{split}
\end{equation}

Since we constructed this classification using MPS ground states, each phase is realized by a system written by the parent Hamiltonian. In Appendix~\ref{ap3}, we showcase the stabilizer Hamiltonians realizing multipole SPT phases with $r=2$ (viz, quadrupole)
in spin chains.

\section{Discussion}\label{sec4}
In accordance with the recent interests in modulated symmetries, we have constructed lattice model of~$(1+1)$d~SPT phases protected by $r$-pole symmetry in the language of MPS.
These phases are classified by the quotients of cohomology groups as indicated in~\eqref{eq: classification of degree-r multipole SPT phases} and~\eqref{eq: classification of degree r multipole in fractional way}. 
Our results deepen the understanding of the infrastructure of modulated symmetries, and in particular, establish a systematic classification of SPT phases with these symmetries.

The results \eqref{eq: classification of degree-r multipole SPT phases} and \eqref{eq: classification of degree r multipole in fractional way} can be generalized to classifications of the other modulated SPTs, where a family of integer-valued polynomials $\ev{f_i^{(r_i)}:\Lambda\to\bbZ\mid \text{degree of }f \text{ is }r_i\in\bbN}$ generates modulated symmetry operators $\prod_{x\in\Lambda}g^{f_i^{(r_i)}(x)}$.
In the main text, we consider an algebra with translation which is generated only by the highest rank multipole symmetry.
In general, however, one needs an algebra with several generators; 
generators may be in different ranks. 
Another question is whether there exists a one-to-one correspondence between our boundary classifications and bulk SPT phases. Equivalently, for each SPT state we obtained on an open chain, can we find the corresponding SPT phase on a closed manifold? Our analysis allowed arbitrary system size~$L$ independent of the finite Abelian symmetry $G$, say $G=\mathbb{Z}_N$. On a closed manifold, however, implementing twisted boundary conditions to select a symmetry-flux sector imposes compatibility constraints between $L$ and $N$. 
The bulk diagnosis in that setting amounts to coupling to a background  gauge fields for multipole symmetries and examining the quantized response~\cite{HanMross2024spinladder}. We defer a systematic treatment of generalizing to other modulated SPT phases and of constraints of twisted boundary conditions and the resulting bulk–boundary correspondence to a separate work.

\par
We conclude this paper with three future directions. 
First, it would be intriguing to extend our systematic classification scheme of the modulated SPT phases to higher dimensions. 
In particular, rigorous classifications of $(2+1)$d multipole SPT phases and semi-injective  PEPS~\cite{MolnarGarreRubioPerezGarciaSchuchCirac1804} can be completed in our approach.
Second, it would also be tempting to establish topological field theory descriptions of the modulated SPT phases. While some classes of modulated SPT states have been discussed from the perspective of anomaly inflow~\cite{anomaly_2024,Lam2311}, full explorations of these theories
have not been made yet. Analyzing these theories, including elucidating the connection to symmetry topological field theories (SymTFTs) would deepen the understanding of the modulated symmetries. Related discussions on subsystem symmetries have being made  in~\cite{Cao:2023rrb,Ohmori:2025fuy,Apruzzi:2025mdl}. 
Third, recent works~\cite{Cao:2023doz,Cao:2024qjj,Pace:2024tgk,Ebisu:2024lie,Pace:2024acq,Seo:2024its,Kim:2025ttl,Pace:2025hpb} study the interplay between noninvertible and modulated symmetries. 
Classifying the SPT phases protected by the combination of noninvertible and modulated symmetries would be another fascinating direction.

\section*{Acknowledgement}
We thank T. Ando, D. Bulmash, M. Honda,
H. Lam, T. Nakanishi, 
S. Pace, K. Shiozaki, L. Li and T. Takama for helpful discussions.
This work is in part supported by JST CREST JPMJCR19T2, JST CREST (Grant Number JPMJCR24I3), 
Villum Fonden Grant no.~VIL60714,  
the Deutsche Forschungsgemeinschaft (DFG, German Research Foundation) under Germany’s Excellence Strategy—Cluster of Excellence Matter
and Light for Quantum Computing (ML4Q) EXC 2004/1 -- 390534769 as well as within the CRC network TR 183 (Project Grant
No. 277101999) as part of subproject B01.
Discussions during the OIST's visiting program TP25QM on ``Generalized Symmetries in Quantum Matter'' were useful to complete this work.


\appendix

\section{Proof of \eqref{mps k}}
\label{sec: proof of k gauge into k+1 gauge}

We give a proof of \cref{mps k}, following \cite{BridgemanChubb1603} and \cite{Lam2311}.

Let $A$ be an injective MPS tensor in left canonical form which forms an SPT ground state \eqref{eq: MPS}.
The left eigenvector of transfer matrix $T(\mathcal{O})=\sum_{h}A^{h\dag}\mathcal{O}A^h$ is identity, and the right eigenvector is a full-rank matrix $\rho$:
\begin{equation}
    \label{eq: eigenequations of transfer matrix T}
    \vcenter{\hbox{
        \input{img/AtransferLeftEE.tex}
    }}
    =
    \vcenter{\hbox{
        \input{img/AtransferLeftEV.tex}
    }}
    ,
    \qquad
    \vcenter{\hbox{
        \input{img/AtransferRightEE.tex}
    }}
    =
    \vcenter{\hbox{
        \input{img/AtransferRightEV.tex}
    }}
    .
\end{equation}
Note that $U_g^{(k+1)}\ket{\psi}$ is a connection of the left-hand side of \eqref{mps k}.
Since $\ket{\psi}$ is a unique ground state of an SPT phase, $\ev{U_g^{(k+1)}}{\psi}$ is a U(1) phase factor.
This value is represented as a connection of a transfer matrix $T_k(\mathcal{O}):=\sum_{h}A^{h\dag}\mathcal{O}X_g^{(k)\dag}A^h$.
If the transfer matrix $T_k$ has a left eigenvalue $|\lambda|\neq 1$, $\ev{U_g^{(k+1)}}{\psi}$ diverges or vanishes in the thermodynamic limit $L\to\infty$.
Thus, $T_k$ has a left eigenvector $X_g^{(k+1)\dag}$ with~U(1)-valued eigenvalue:
\begin{equation}
    \label{eq: left eigenequation of transfer matrix Tk}
    \vcenter{\hbox{
        \input{img/AXtransfLEE.tex}
    }}
    \ueq
    \vcenter{\hbox{
        \input{img/AXtransfLEV.tex}
    }}
    .
\end{equation}
Now we evaluate the following quantity:
\begin{equation}
    \label{eq: XXrho is half splitted}
    \begin{split}
        \vcenter{\hbox{
            \input{img/XdagXrho.tex}
        }}
        &=
        \vcenter{\hbox{
            \input{img/XXAXrhoA.tex}
        }}
        \\
        &
        =
        \vcenter{\hbox{
            \input{img/XAXrhorhoAX.tex}
        }}
        .
    \end{split}
\end{equation}
\eqref{eq: eigenequations of transfer matrix T} derives that the upper and lower halves of the right-hand side have an equal norm as the following:
\begin{equation}
    \begin{split}
        \vcenter{\hbox{
            \input{img/Xk1Anorm.tex}
        }}
        &=
        \vcenter{\hbox{
            \input{img/Xk1norm.tex}
        }}
        \\
        &=
        \vcenter{\hbox{
            \input{img/XAXnorm.tex}
        }}
        .
    \end{split}
\end{equation}
In the last equality, we used the unitarity of $X_g^{(k)}$.
Compared with \eqref{eq: XXrho is half splitted}, we have
\begin{equation}
    \begin{split}
        &
        \qty|
            \vcenter{\hbox{
                \input{img/XAXrhorhoAX.tex}
            }}
        |^2
        \\
        &=
        \vcenter{\hbox{
            \input{img/Xk1Anorm.tex}
        }}
        \vcenter{\hbox{
            \input{img/XAXnorm.tex}
        }}
        ,
    \end{split}
\end{equation}
which is nothing but the Cauchy-Schwarz equality; the upper and lower half of the left-hand side, which have the same norm, are proportional to each other.
Therefore, 
\begin{equation}
    \vcenter{\hbox{
        \input{img/XArhovec.tex}
    }}
    \ueq
    \vcenter{\hbox{
        \input{img/XAXrhovec.tex}
    }}
\end{equation}
which is equivalent to \eqref{mps k} if $X_g^{(k+1)}$ is a unitary operator.

From \eqref{mps k} and \eqref{eq: left eigenequation of transfer matrix Tk}, it follows that
\begin{equation}
    \begin{split}
        \vcenter{\hbox{
            \input{img/Xk1AAXk1vec.tex}
        }}
        &\ueq
        \vcenter{\hbox{
            \input{img/Xk1AXk1AXkvec.tex}
        }}
        \\
        &\ueq
        \vcenter{\hbox{
            \input{img/XXvec.tex}
        }}
        .
    \end{split}
\end{equation}
U(1)-valued eigenvalue of a transfer matrix built from an injective MPS tensor is always 1, and the corresponding left eigenvector is always identity.
Therefore, $X_g^{(k+1)}$ is exactly a unitary operator.

\section{Proof of \cref{prop:simplified commutativity condition}}
\label{sec: proof of prop. simplified commutativity condition}
In this section, we prove the \cref{prop:simplified commutativity condition}, which includes both directions between \cref{item: commutativity condition in rough form} and \cref{item: commutativity condition in simplified form}. Before delving into the proof, we introduce several conventions.
We prove the proposition for fixed $g,h\in G$, which can be straightforwardly generalized to arbitrary group elements. For simplicity, we omit the group element subscripts of $\theta_{g,h}(k,l)$.
With
\begin{equation}
    S_L(k,l)
    :=
    \sum_{m=0}^{k}
    \sum_{n=0}^{l}
    \theta(m,n)
    B_L(k-m)B_L(l-n),\quad B_L(t)
    :=
    \binom{L+t-1}{t},
\end{equation}
and
\begin{equation}
    C_L(k,l)
    :=
    S_L(k,l)-\theta(k,l),
\end{equation}
the identity \eqref{eq: commutativity condition in rough form} is rewritten as
\begin{equation}
    \exp[iC_L(k,l)]=1.
\end{equation}
For a given function $f:\bbZ^2\to\bbR$, we define the Laplacian
\begin{equation}
    \Delta f(k,l)
    :=
    f(k,l)-f(k-1,l)-f(k,l-1)+f(k-1,l-1)
    .
\end{equation}
Further with 
\begin{equation}
    \label{eq: Pascal identity for L-1 t and L t-1}
    B_L(t)
    =
    B_{L}(t-1)
    +
    B_{L-1}(t),
\end{equation}
from rewriting Pascal's identity
\begin{equation}
    \binom{L+t-1}{t}
    =
    \binom{L+t-2}{t-1}
    +
    \binom{L+t-2}{t},
\end{equation}
we can simplify $\Delta S_L(k,l)$ as
\begin{equation}
    \label{eq: Delta S_L is S_L-1}
    \Delta S_L(k,l)
    =
    S_{L-1}(k,l)
    .
\end{equation}
The computation is the following
\begin{equation}
    \begin{split}
        &
        \Delta S_L(k,l) 
        \\
        = &
        \sum_{m=0}^{k-1}
        \sum_{n=0}^{l-1}
        \theta(m,n)
        (
            B_{L}(k-m)B_{L}(l-n)
            -
            B_{L}(k-1-m)B_{L}(l-n)
            \\
            &-
            B_{L}(k-m)B_{L}(l-1-n)
            +
            B_{L}(k-1-m)B_{L}(l-1-n)
        )
        \\
        + &
        \qty(
            \sum_{m=0}^{k-1}B_{L}(k-m)\delta_{n,l}
            +
            \sum_{n=0}^{l-1}\delta_{m,k}B_{L}(l-n)
            +
            \delta_{k,m}\delta_{l,n}
            -
            \sum_{m=0}^{k-1}B_{L}(k-1-m)\delta_{n,l}
            -
            \sum_{n=0}^{l-1}\delta_{m,k}B_{L}(l-1-n)
        )
        \theta(m,n)
        \\
        =&
        \sum_{m=0}^{k-1}
        \sum_{n=0}^{l-1}
        \theta(m,n)
        \qty(B_L(k-m)-B_L(k-1-m))
        \qty(B_L(l-n)-B_L(l-1-n))
        \\
        &\qquad
        +
        \sum_{m=0}^{k-1}
        \theta(m,l)
        \qty(B_L(k-m)-B_L(k-1-m))
        \\
        &\qquad
        +
        \sum_{n=0}^{l-1}
        \theta(k,n)
        \qty(B_L(l-n)-B_L(l-1-n))
        +
        \theta(k,l)
        \\
        =&
        \sum_{m=0}^{k-1}
        \sum_{n=0}^{l-1}
        \theta(m,n)
        B_{L-1}(k-m)B_{L-1}(l-n)
        +
        \sum_{m=0}^{k-1}
        \theta(m,l)
        B_{L-1}(k-m)
        \\
        &\qquad
        +
        \sum_{n=0}^{l-1}
        \theta(k,n)
        B_{L-1}(l-n)
        +
        \theta(k,l),
    \end{split}
\end{equation}
where we used \eqref{eq: Pascal identity for L-1 t and L t-1} in the last line.

\subsection{\cref{item: commutativity condition in rough form} $\Rightarrow$ \cref{item: commutativity condition in simplified form}}

First we prove sufficiency (\cref{item: commutativity condition in rough form} $\Rightarrow$ \cref{item: commutativity condition in simplified form}).


From \cref{item: commutativity condition in rough form}, $C_L(k,l)=0$ for all $L$ and $k,l\leq r$. Therefore,
\begin{equation}
    \begin{split}
        0
        &=
        \Delta C_L(k,l)
        =
        \Delta S_L(k,l)
        -
        \Delta \theta(k,l)
        \\
        &=
        S_{L-1}(k,l)
        -
        \Delta \theta(k,l)
        \\
        &=
        \theta(k-1,l)-\theta(k,l-1)+\theta(k-1,l-1),
    \end{split}
\end{equation}
where we used \eqref{eq: Delta S_L is S_L-1} in the second line, and  $C_{L-1}=S_{L-1}-\theta=0$ in the last equality.
This is equivalent to the second equation in \eqref{eq: commutativity condition in simplified form}.


We give a proof of $\theta(k,l)=0$ for $k+l<r$ by induction.
$C_L(0,1)=\theta(0,0)L=0$ guarantees $\theta(0,0)=0$, which is the case of $k+l=0$.
Let us assume $\theta(k,l)=0$ for all $k+l<s<r$.
Since we have proven $\theta(k,l)=\theta(k+1,l)+\theta(k,l+1)$, $\theta(0,s+1)$ determines the other $\theta(k,l)$ for $k+l=s$:
\begin{equation}
    \theta(0,s)
    =
    -\theta(1,s-1)
    =
    \theta(2,s-2)
    =
    \cdots
    =
    (-1)^{s}\theta(s,0)
\end{equation}
On the other hand,
\begin{equation}
    \begin{split}
        0
        &=
        C_L(0,s+1)
        =
        \sum_{n=0}^{s}
        \theta(0,n)
        B_L(s+1-n)
        =
        L\theta(0,s)
        ,
    \end{split}
\end{equation}
therefore $\theta(k,l)=0$ for all $k+l=s$.

\subsection{\cref{item: commutativity condition in rough form} $\Leftarrow$ \cref{item: commutativity condition in simplified form}}

Now we assume \cref{item: commutativity condition in simplified form}, which is $\theta(k,l)=\theta(k+1,l)+\theta(k,l+1)$ for $k,l<r$ and $\theta(k,l)=0$ for $k+l<r$.
We can rewrite the former equation as
\begin{equation}
    \label{eq: Delta theta is theta}
    \Delta \theta(k,l)=\theta(k,l),\quad \forall k,l<r+1.
\end{equation}


By \eqref{eq: Delta S_L is S_L-1} and \eqref{eq: Delta theta is theta}, 
\begin{equation}
    \Delta C_L(k,l)
    =
    \Delta S_L(k,l)-\Delta \theta(k,l)
    =
    S_{L-1}(k,l)-\theta(k,l)
    .
\end{equation}
Thus we have
\begin{equation}
    \label{eq: Delta C_L is C_L-1}
    \Delta C_L(k,l)
    =
    C_{L-1}(k,l)
    .
\end{equation}


Although it is not physical to consider $L=1$, $C_1$ is well-defined mathematically:
\begin{equation}
    C_1(k,l)
    =
    \sum_{m=0}^{k}
    \sum_{n=0}^{l}
    \theta(m,n)
    B_1(k-m)B_1(l-n)
    -
    \theta(k,l)
    =
    \sum_{m=0}^{k}
    \sum_{n=0}^{l}
    \theta(m,n)
    -
    \theta(k,l).
\end{equation}
When $k+l\leq r$, the right hand side is trivially zero.
When $k+l>r$, we get
\begin{equation}
    \begin{split}
        \Delta C_1(k,l)
        &=
        \qty(
            \sum_{m=0}^{k}
            \sum_{n=0}^{l}
            -
            \sum_{m=0}^{k-1}
            \sum_{n=0}^{l}
            -
            \sum_{m=0}^{k}
            \sum_{n=0}^{l-1}
            +
            \sum_{m=0}^{k-1}
            \sum_{n=0}^{l-1}
        )
        \theta(m,n)
        -
        \Delta\theta(k,l)
        \\
        &=
        \theta(k,l)-\theta(k,l)
        =0
        .
    \end{split}
\end{equation}
Therefore $C_1(k,l)=0$ for all $k,l\geq 0$.


We prove $C_L(k,l)=0$ for all $L\geq 1$ and $k,l\geq 0$ by induction of $L$.
Now we assume $C_{L-1}(k,l)=0$ for all $k,l\geq 0$.
By \eqref{eq: Delta C_L is C_L-1}, 
\begin{equation}
    C_L(k,l)
    =
    C_L(k-1,l)
    +
    C_L(k,l-1)
    -
    C_L(k-1,l-1)
\end{equation}
i.e. $C_L(k,l)=0$ if $C_L(m,n)=0$ for all $m+n<k+l$.
If $m+n\leq r$, 
\begin{equation}
    C_L(m,n)
    =
    \sum_{m'=0}^{m}
    \sum_{n'=0}^{n}
    \theta(m',n')
    B_L(m-m')B_L(n-n')
    -
    \theta(m,n)
    =
    0
\end{equation}
because surviving $\theta(m',n')$ is zero by assumption.
Therefore by induction of $k+l$, we have $C_L(k,l)=0$ for all $k,l\geq 0$.
This is equivalent to \eqref{eq: commutativity condition in rough form}.

\section{Lattice realizations of quadrupole SPT states in distinct classes}\label{ap3}
In this section, we demonstrate the classification of quadrupole ($r=2$) SPTs through concrete spin models.
Construction strategy of models is based on \cite{HanLakeLamVerresenYou2309}.
According to our formula~\eqref{eq: classification of degree-r multipole SPT phases}, 
the SPT phases are classified by
\begin{eqnarray}
    [H^2(G,U(1))]^{\oplus 2}
            \oplus 
            \qty[\frac{H^1(G,H^1(G,U(1)))}{H^2(G,U(1))}],
\end{eqnarray}
with three different projective representations on the edge described by 
\begin{align}
    \begin{split}
       \text{monopole-quadrupole:}~~ &e^{i\theta_{g,h}(1,1)}=e^{i(\theta_{g,h}(1,2)-\theta_{h,g}(1,2))}=e^{-i\theta_{h,g}(0,2)},\\
       \text{dipole-quadrupole:}~~ &e^{i(\theta_{g,h}(1,2)+\theta_{h,g}(1,2))},\\
       \text{quadrupole-quadrupole:}~~ &e^{i\theta_{g,h}(2,2)}.
    \end{split}
\end{align}
In the following, we give three classes of SPT protected by quadrupole symmetry realizing each case. 
\subsection{The first class -- monopole-quadrupole SPTs}
We think of a spin chain with $\mathbb Z_N\times \mathbb Z_M$ spin on each site. The clock and shift operators satisfy the following commutation relations
\begin{align}
    Z_{j}X_{j}=e^{\frac{2\pi i}{N}}X_{j}Z_{j},\quad \tilde{Z}_j\tilde{X}_j=e^{\frac{2\pi i}{M}}\tilde{X}_j\tilde{Z}_j.
\end{align}
Assume we have $\mathbb Z_N\times \mathbb Z_M$ quadrupole symmetry, together with $\mathbb Z_N\times \mathbb Z_M$ dipole and charge symmetry The symmetry operators are
\begin{align}\label{eq:qsymop1}
    \begin{split}
        &Q=\prod_{j}X_j,\quad D=\prod_{j}X_j^j,\quad \text{Qu}=\prod_{j}X_j^{j(j+1)/2},\\
        &\tilde{Q}=\prod_{j}\tilde{X}_j,\quad \tilde{D}=\prod_{j}\tilde{X}_j^j,\quad \tilde{\text{Qu}}=\prod_{j}\tilde{X}_j^{j(j+1)/2}.
    \end{split}
\end{align}
Consider quadrupole symmetric stabilizer  Hamiltonian
\begin{equation}\label{eq:quspt1}
    H=-\sum_{j}a_j+\tilde{a}_j+\text{h.c.},
\end{equation}
where
\begin{equation}
    a_j=(\tilde{Z}_{j-1}\tilde{Z}_{j}^{3\dagger})^{\frac{M\xi}{K}}X_{j+1}^{\dagger}(\tilde{Z}_{j+1}^{3}\tilde{Z}_{j+2}^{\dagger})^{\frac{M\xi}{K}},\quad \tilde{a}_j=(Z_{j-1}Z_{j}^{3\dagger})^{\frac{N\xi}{K}}\tilde{X}_{j}(Z_{j+1}^3Z_{j+2}^{\dagger})^{\frac{N\xi}{K}}.
\end{equation}
The parameter $\xi$ takes value in $\mathbb Z_K$ and $K=\text{gcd}(M,N)$. The model \eqref{eq:quspt1} contains terms with single-site charge operator $X_{j+1}$ ($\tilde{X}_j$) decorated by the quadrupole operator $(\tilde{Z}_{j-1}\tilde{Z}_{j}^{3\dagger}\tilde{Z}_{j+1}^{3}\tilde{Z}_{j+2}^{\dagger})^{\dagger\frac{M\xi}{K}}$ ($(Z_{j-1}Z_{j}^{3\dagger}Z_{j+1}^3Z_{j+2}^{\dagger})^{\frac{N\xi}{K}}$), and therefore realizes SPT protected by $\mathbb Z_N\times \mathbb Z_M$ quadrupole symmetry. In Ref.~\cite{Lam2311}, ignoring the quadrupole symmetry, this Hamiltonian is used to study the classification of dipole SPT. Here we show the full projective representations at the boundaries and relate this model with the MPS classification for quadrupole symmetry.

For open boundary condition, the symmetry operators~\eqref{eq:qsymop1} factorize at each boundary. It is enough to focus on the part at the left boundary near site $1$
\begin{align}
    \begin{split}
        &\mathcal L_{Q}=X_1X_2(\tilde{Z}_{1}\tilde{Z}_{2}^{2\dagger}\tilde{Z}_{3})^{\frac{M\xi}{K}},\quad \mathcal L_{\tilde{Q}}=\tilde{X}_1(Z_{1}^{\dagger}Z_{2}^{2}Z_{3}^{\dagger})^{\frac{N\xi}{K}},\\
        &\mathcal L_{D}=X_1X_2^2(\tilde{Z}_{1}^{3}\tilde{Z}_{2}^{5\dagger}\tilde{Z}_{3}^{2})^{\frac{M\xi}{K}},\quad \mathcal L_{\tilde{D}}=\tilde{X}_1(Z_{1}^{2\dagger}Z_{2}^{3}Z_{3}^{\dagger})^{\frac{N\xi}{K}}\\
        &\mathcal L_{\text{Qu}}=X_1X_2^3(\tilde{Z}_{1}^{6}\tilde{Z}_{2}^{8\dagger}\tilde{Z}_{3}^{3})^{\frac{M\xi}{K}},\quad \mathcal L_{\tilde{\text{Qu}}}=\tilde{X}_1(Z_{1}^{3\dagger}Z_{2}^{3}Z_{3})^{\frac{N\xi}{K}},
    \end{split}
\end{align}
which gives the projective representation between charge, dipole and quadrupole symmetry
\begin{align}
    \begin{split}
        &\mathcal L_{D}\mathcal L_{\tilde{D}}=e^{-\frac{2\pi i\xi}{K}}\mathcal L_{\tilde{D}}\mathcal L_{D},\quad \mathcal L_{\text{Qu}}\mathcal L_{\tilde{D}}=e^{-\frac{2\pi i\xi}{K}}\mathcal L_{\tilde{D}}\mathcal L_{\text{Qu}}\\
        &\mathcal L_{Q}\mathcal L_{\tilde{\text{Qu}}}=e^{\frac{2\pi i\xi}{K}}L_{\tilde{\text{Qu}}}\mathcal L_{Q},\quad \mathcal L_{\text{Qu}}\mathcal L_{\tilde{Q}}=e^{\frac{2\pi i\xi}{K}}L_{\tilde{Q}}\mathcal L_{\text{Qu}}.
    \end{split}
\end{align}
Therefore, this model corresponds to the MPS class
\begin{equation}
    e^{i\theta_{g,h}(1,1)}=e^{i(\theta_{g,h}(1,2)-\theta_{h,g}(1,2))}=e^{-i\theta_{h,g}(0,2)},
\end{equation}
with $H^2(G,U(1))$ classification. This is consistent with \eqref{eq:quspt1} labelled by $\xi\in\mathbb Z_{\text{gcd}(M,N)}=H^2(\mathbb Z_M\times \mathbb Z_N,U(1))$. 
For general abelian group $G=\prod_{I}\mathbb Z_{I}$, only the decorations of quadrupole operators on single-site charge operator from a different cyclic factor in $G$ commute with each other. Therefore, the classification is
\begin{equation}
    \prod_{I<J}\mathbb Z_{IJ}=H^2(G,U(1)),\quad  \mathbb Z_{IJ}=\mathbb Z_{\text{gcd}(I,J)}.
\end{equation}
Following the similar logic, we give the other two classes of quadrupole SPT Hamiltonians (and their classifications).

\subsection{The second class -- dipole-quadrupole SPTs}
Consider the following SPT Hamiltonian (with octupole symmetry) on a open chain with $\mathbb Z_N$ spins
\begin{equation}
    H=-\sum_{j=3}^{L-2}(Z_{j-2}Z_{j-1}^{4\dagger}Z_{j}^3)^{k}X_{j}(Z_{j}^3Z_{j+1}^{4\dagger}Z_{j+2})^{k}+h.c.\quad k=0,...,N-1.\label{c2}
\end{equation}
Note that the model~\eqref{c2} respects 
higher rank multipole symmetry than the quadrupole, 
that is, octupole symmetry. 
Here, we allow continuous perturbation which violates octupole symmetry yet respects 
lower rank symmetries, i.e., 
quadrupole, dipole, and global symmetries. See also Ref.~\cite{Lam2311} for the relevant discussion in the case of the SPT phase with dipole symmetry. 
Acting on the ground state, the symmetry operators factorize at each boundary. At the left boundary, we have
\begin{align}
    \begin{split}
        &\mathcal L_{Q}=X_1X_2(Z_{1}^{\dagger}Z_{2}^{3}Z_{3}^{3\dagger}Z_{4})^{k}, \\
        &\mathcal L_{D} =X_1X_2^2(Z_{1}^{3\dagger}Z_{2}^{8}Z_{3}^{7\dagger}Z_{4}^{2})^{k}\\
        &\mathcal L_{Qu}=X_1X_2^3(Z_{1}^{6\dagger}Z_{2}^{14}Z_{3}^{11\dagger}Z_{4}^{3})^{k}
    \end{split}
\end{align}
Then we have the only nontrivial commutation relation between dipole and quadrupole symmetry operators
\begin{align}
    \begin{split}
    \mathcal L_{D}\mathcal L_{Qu}=e^{-\frac{2\pi ik}{N}} \mathcal L_{Qu}\mathcal L_{D},
    \end{split}
\end{align}
Therefore, this model corresponds to the MPS class characterized by 
\begin{equation}
    e^{i(\theta_{g,h}(1,2)+\theta_{h,g}(1,2))} \in \frac{H^1(G,H^1(G,U(1)))}{H^2(G,U(1))}. 
\end{equation}

\subsection{The third class -- quadrupole-quadrupole SPTs}

Consider the following SPT (with Hexadecapole symmetry)
\begin{equation}
   H=-\sum_{j=3}^{L-3}a_j+\tilde{a}_j+h.c.,
\end{equation}
where
\begin{align}
    \begin{split}
        a_j=(\tilde{Z}_{j-2}\tilde{Z}_{j-1}^{5\dagger}\tilde{Z}_{j}^{10})^{\frac{M\xi}{K}}X_{j+1}^{\dagger}(\tilde{Z}_{j+1}^{10\dagger}\tilde{Z}_{j+2}^{5}\tilde{Z}_{j+3}^{\dagger})^{\frac{M\xi}{K}},\quad \tilde{a}_j=(Z_{j-2}Z_{j-1}^{5\dagger}Z_{j}^{10})^{\frac{N\xi}{K}}\tilde{X}_{j}(Z_{j+1}^{10\dagger}Z_{j+2}^{5}Z_{j+3}^{\dagger})^{\frac{N\xi}{K}}.\label{c3}
    \end{split}
\end{align}
{Similar to the previous section, the model~\eqref{c3} has higher rank multipole symmetries than quadrupole, which are octupole and hexadecapole symmetries. However, we think of adding perturbations so that the model keeps global, dipole, and quadrupole symmetries. }\par
On the left boundary
\begin{align}
    \begin{split}
        &\mathcal L_{Q}=X_1X_2X_3(\tilde{Z}_1\tilde{Z}_2^{4\dagger}\tilde{Z}_3^6\tilde{Z}_{4}^{4\dagger}\tilde{Z}_{5})^{\frac{M\xi}{K}},\quad \mathcal L_{\tilde{Q}}=\tilde{X}_1\tilde{X}_2(Z_1^{\dagger}Z_2^{4}Z_3^{6\dagger}Z_{4}^{4}Z_5^{5\dagger})^{\frac{N\xi}{K}}\\
        &\mathcal L_{D}=X_1X_2^2X_3^3(\tilde{Z}_1^{4}\tilde{Z}_2^{15\dagger}\tilde{Z}_3^{21}\tilde{Z}_{4}^{13\dagger}\tilde{Z}_{5}^3)^{\frac{M\xi}{K}},\quad \mathcal L_{\tilde{D}}=\tilde{X}_1\tilde{X}_2^2(Z_1^{3\dagger}Z_2^{11}Z_3^{15\dagger}Z_{4}^{9}Z_{5}^{2\dagger})^{\frac{N\xi}{K}}\\
        &\mathcal L_{Qu}=X_1X_2^3X_3^6(\tilde{Z}_{1}^{10}\tilde{Z}_{2}^{35\dagger}\tilde{Z}_{3}^{46}\tilde{Z}_{4}^{27\dagger}\tilde{Z}_{5}^{6})^{\frac{M\xi}{K}},\quad \mathcal L_{\tilde{Qu}}=\tilde{X}_1\tilde{X}_2^3(Z_1^{6\dagger}Z_2^{20}Z_3^{25\dagger}Z_{4}^{14}Z_{5}^{3\dagger})^{\frac{N\xi}{K}},
    \end{split}
\end{align}
with the only nontrivial commutation relation between quadrupole symmetry opreators
\begin{align}
    \begin{split}
    \mathcal L_{Qu}\mathcal L_{\tilde{Qu}}=e^{\frac{2\pi i\xi}{K}}\mathcal L_{\tilde{Qu}}\mathcal L_{Qu}
    \end{split}
\end{align}
Therefore, this class corresponds to the MPS class characterized by 
\begin{equation}
    e^{i\theta_{g,h}(2,2)} \in  H^2(G,U(1)).
\end{equation}

\bibliography{main}
\bibliographystyle{ytphys}

\end{document}

%% file: img/MPS.tex
\begin{tikzpicture}
    \foreach\x in{0,1}{
        \node at(\x,0)[draw,rectangle](A\x){$A$};
        \draw[ultra thick](A\x.west)--++(-.45,0);
        \draw[ultra thick](A\x.east)--++(.45,0);
        \draw(A\x.north)--++(0,.5);
    }
\end{tikzpicture}

%% file: img/monopole-MPS.tex
\begin{tikzpicture}
    \node at(0,0)[draw,rectangle](A){$A$};
    \node at(0,.75)[draw,rectangle](U){$g$};
    \draw[ultra thick](A.west)--++(-.5,0);
    \draw[ultra thick](A.east)--++(.5,0);
    \draw(A.north)--(U.south);
    \draw(U.north)--++(0,.2);
\end{tikzpicture}

%% file: img/monopole-act-result.tex
\begin{tikzpicture}
    \node at(.1,0)[draw,rectangle](A){$A$};
    \node at(1,0)[draw,rectangle](E){$X_g^{(0)}$};
    \node at(-1,0)[draw,rectangle](W){$X_g^{(0)\dag}$};
    \draw[ultra thick](W.west)--++(-.2,0);
    \draw[ultra thick](E.east)--++(.2,0);
    \draw[ultra thick](A.west)--(W.east);
    \draw[ultra thick](A.east)--(E.west);
    \draw(A.north)--++(0,.2);
\end{tikzpicture}

%% file: img/dipole-MPS.tex
\begin{tikzpicture}
    \node at(0,0)[draw,rectangle](A1){$A^{(x)}$};
    \node at(0,.75)[draw,rectangle](U1){$g_{x}$};
    \node at(2,0)[draw,rectangle](A2){$A^{(x+1)}$};
    \node at(2,.75)[draw,rectangle](U2){$g_{x+1}$};
    \foreach\x in{1,2}{
        \draw (A\x.north) -- (U\x.south);
        \draw (U\x.north) -- ++(0,.2);
        \draw[ultra thick] (A\x.east) -- ($(A\x)+(1,0)$);
        \draw[ultra thick] (A\x.west) -- ($(A\x)+(-1,0)$);
    }
\end{tikzpicture}

%% file: img/mono-gauge.tex
\begin{tikzpicture}
    \node at(-1.,0)[draw,rectangle](W0){\tiny $X_g^{(0)\dag}$};
    \node at(.5,0)[draw,rectangle](A0){$A^{(x)}$};
    \draw[ultra thick](W0.west)--++(-.3,0);
    \draw[ultra thick](W0.east)--(A0.west);
    \draw[ultra thick](A0.east)--++(.5,0);
    \draw(A0.north)--++(0,.3);
\end{tikzpicture}

%% file: img/dipole-newgauge.tex
\begin{tikzpicture}
    \node at(-1.1,0)[draw,rectangle](W0){\tiny $X_g^{(1)\dag}$};
    \node at(0,0)[draw,rectangle](A0){$A^{(x)}$};
    \node at(1.,0)[draw,rectangle](E0){\tiny $X_g^{(1)}$};
    \draw[ultra thick](W0.west)--++(-.3,0);
    \draw[ultra thick](W0.east)--(A0.west);
    \draw[ultra thick](A0.east)--(E0.west);
    \draw[ultra thick](E0.east)--++(.5,0);
    \draw(A0.north)--++(0,.3);
\end{tikzpicture}

%% file: img/k-gauge.tex
\begin{tikzpicture}
    \node at(-1.,0)[draw,rectangle](W0){\tiny $X_g^{(k)\dag}$};
    \node at(.5,0)[draw,rectangle](A0){$A^{(x)}$};
    \draw[ultra thick](W0.west)--++(-.3,0);
    \draw[ultra thick](W0.east)--(A0.west);
    \draw[ultra thick](A0.east)--++(.5,0);
    \draw(A0.north)--++(0,.3);
\end{tikzpicture}

%% file: img/k-new-gauge.tex
\begin{tikzpicture}
    \node at(-1.2,0)[draw,rectangle](W0){\tiny $X_g^{(k+1)\dag}$};
    \node at(0,0)[draw,rectangle](A0){$A^{(x)}$};
    \node at(1.2,0)[draw,rectangle](E0){\tiny $X_g^{(k+1)}$};
    \draw[ultra thick](W0.west)--++(-.2,0);
    \draw[ultra thick](W0.east)--(A0.west);
    \draw[ultra thick](A0.east)--(E0.west);
    \draw[ultra thick](E0.east)--++(.2,0);
    \draw(A0.north)--++(0,.3);
\end{tikzpicture}

%% file: img/AtransferLeftEE.tex
\begin{tikzpicture}
    \node at(0,0)[draw,rectangle](A1){$A$};
    \node at(0,-1)[draw,rectangle](A2){$A$};
    \draw[ultra thick](A1.west)--++(-.5,0)|-(A2.west);
    \draw[ultra thick](A1.east)--++(.5,0);
    \draw[ultra thick](A2.east)--++(.5,0);
    \draw(A1.south)--(A2.north);
\end{tikzpicture}

%% file: img/AtransferLeftEV.tex
\begin{tikzpicture}
    \draw[ultra thick](0,0)--++(-1,0)|-(0,-1);
\end{tikzpicture}

%% file: img/AtransferRightEE.tex
\begin{tikzpicture}
    \node at(0,0)[draw,rectangle](A1){$A$};
    \node at(0,-1)[draw,rectangle](A2){$A$};
    \node at(.75,-.5)[draw,rectangle](rho){$\rho$};
    \draw[ultra thick](A1.east)-|(rho.north);
    \draw[ultra thick](A2.east)-|(rho.south);
    \draw[ultra thick](A1.west)--++(-.5,0);
    \draw[ultra thick](A2.west)--++(-.5,0);
    \draw(A1.south)--(A2.north);
\end{tikzpicture}

%% file: img/AtransferRightEV.tex
\begin{tikzpicture}
    \node at(.75,-.5)[draw,rectangle](rho){$\rho$};
    \draw[ultra thick](0,0)-|(rho.north);
    \draw[ultra thick](0,-1)-|(rho.south);
\end{tikzpicture}

%% file: img/AXtransfLEE.tex
\begin{tikzpicture}
    \node at(0,0)[draw,rectangle](A1){$A$};
    \node at(0,-1)[draw,rectangle](A2){$A$};
    \node at(-1,-1)[draw,rectangle](Xk){$X^{(k)\dag}_g$};
    \node at(-2.5,-.5)[draw,rectangle](Xk2){$X^{(k+1)}_g$};
    \draw[ultra thick](A1.west)-|(Xk2.north);
    \draw[ultra thick](A2.west)--(Xk.east);
    \draw[ultra thick](Xk.west)-|(Xk2.south);
    \draw[ultra thick](A1.east)--++(.5,0);
    \draw[ultra thick](A2.east)--++(.5,0);
    \draw(A1.south)--(A2.north);
\end{tikzpicture}

%% file: img/AXtransfLEV.tex
\begin{tikzpicture}
    \node at(-.5,-.5)[draw,rectangle](Xk2){$X^{(k+1)}_g$};
    \draw[ultra thick](Xk2.north)|-(.5,0);
    \draw[ultra thick](Xk2.south)|-(.5,-1);
\end{tikzpicture}

%% file: img/XdagXrho.tex
\begin{tikzpicture}
    \node at(1.5,0)[draw,rectangle](Xkdag){\tiny $X^{(k+1)\dag}$};
    \node at(0,0)[draw,rectangle](Xk){\tiny $X^{(k+1)}$};
    \node at(2.5,.5)[draw,rectangle](rho){$\rho$};
    \coordinate (NW) at (-1,1);
    \draw[ultra thick]
        (Xk.east)--(Xkdag.west)
        (Xkdag.east)-|(rho.south)
        (rho.north)|-(NW)
        (NW)|-(Xk.west)
    ;
\end{tikzpicture}

%% file: img/XXAXrhoA.tex
\begin{tikzpicture}
    \node at(1.5,0)[draw,rectangle](Xk1dag){$X^{(k+1)\dag}$};
    \node at(0,0)[draw,rectangle](A){$A$};
    \node at(0,1)[draw,rectangle](A2){$A$};
    \node at(-2.5,0)[draw,rectangle](Xk1){$X^{(k+1)}$};
    \node at(-1,0)[draw,rectangle](Xkdag){$X^{(k)\dag}$};
    \node at(2.5,.5)[draw,rectangle](rho){$\rho$};
    \coordinate (NW) at (-3.5,1);
    \draw[ultra thick]
        (Xk1.east)--(Xkdag.west)
        (Xkdag.east)--(A.west)
        (A.east)--(Xk1dag.west)
        (Xk1dag.east)-|(rho.south)
        (rho.north)|-(A2.east)
        (A2.west)--(NW)
        (NW)|-(Xk1.west)
    ;
    \draw(A2.south)--(A.north);
\end{tikzpicture}

%% file: img/XAXrhorhoAX.tex
\begin{tikzpicture}
    \node at(0,0)[draw,rectangle](A){$A$};
    \node at(0,2)[draw,rectangle](A2){$A$};
    \node at(-1.25,2)[draw,rectangle](Xk1){$X^{(k+1)}$};
    \node at(1.25,0)[draw,rectangle](Xk1dag){$X^{(k+1)\dag}$};
    \node at(-1,0)[draw,rectangle](Xkdag){$X^{(k)\dag}$};
    \node at(2.5,.5)[draw,rectangle](rhoS){$\sqrt{\rho}$};
    \node at(2.5,1.5)[draw,rectangle](rhoN){$\sqrt{\rho}$};
    \coordinate (NW) at (-2.25,2);
    \draw[ultra thick]
        (Xk1.west)--(NW)|-(Xkdag.west)
        (Xkdag.east)--(A.west)
        (A.east)--(Xk1dag.west)
        (Xk1dag.east)-|(rhoS.south)
        (rhoS.north)--(rhoN.south)
        (rhoN.north)|-(A2.east)
        (A2.west)--(Xk1.east)
    ;
    \draw(A2.south)--(A.north);
    \draw[dashed](-2.5,1)--(3,1);
\end{tikzpicture}

%% file: img/Xk1Anorm.tex
\begin{tikzpicture}
    \node at(0,0)[draw,rectangle](AS){$A$};
    \node at(0,2)[draw,rectangle](AN){$A$};
    \node at(-1.,0)[draw,rectangle](Xk1dag){\tiny $X^{(k+1)\dag}$};
    \node at(-1.,2)[draw,rectangle](Xk1){\tiny $X^{(k+1)}$};
    \node at(0.75,0.5)[draw,rectangle](rhoS){$\sqrt{\rho}$};
    \node at(0.75,1.5)[draw,rectangle](rhoN){$\sqrt{\rho}$};
    \coordinate (NW) at (-2,2);
    \draw(AS.north)--(AN.south);
    \draw[ultra thick]
        (AS.east)-|(rhoS.south)
        (rhoS.north)--(rhoN.south)
        (rhoN.north)|-(AN.east)
        (AN.west)--(Xk1.east)
        (Xk1.west)--(NW)|-(Xk1dag.west)
        (Xk1dag.east)--(AS.west)
    ;
\end{tikzpicture}

%% file: img/Xk1norm.tex
\begin{tikzpicture}
    \node at(-0.5,0)[draw,rectangle](Xk1dag){$X^{(k+1)\dag}$};
    \node at(-0.5,2)[draw,rectangle](Xk1){$X^{(k+1)}$};
    \node at(0.75,0.5)[draw,rectangle](rhoS){$\sqrt{\rho}$};
    \node at(0.75,1.5)[draw,rectangle](rhoN){$\sqrt{\rho}$};
    \coordinate (NW) at (-1.5,2);
    \draw[ultra thick]
        (Xk1dag.east)-|(rhoS.south)
        (rhoS.north)--(rhoN.south)
        (rhoN.north)|-(Xk1.east)
        (Xk1.west)--(NW)|-(Xk1dag.west)
    ;
\end{tikzpicture}

%% file: img/XAXnorm.tex
\begin{tikzpicture}
    \node at(0,0)[draw,rectangle](A){$A$};
    \node at(0,2)[draw,rectangle](A2){$A$};
    \node at(1.25,2)[draw,rectangle](Xk1){$X^{(k+1)}$};
    \node at(-1,2)[draw,rectangle](Xk){$X^{(k)}$};
    \node at(1.25,0)[draw,rectangle](Xk1dag){$X^{(k+1)\dag}$};
    \node at(-1,0)[draw,rectangle](Xkdag){$X^{(k)\dag}$};
    \node at(2.5,.5)[draw,rectangle](rhoS){$\sqrt{\rho}$};
    \node at(2.5,1.5)[draw,rectangle](rhoN){$\sqrt{\rho}$};
    \coordinate (NW) at (-2.,2);
    \draw[ultra thick]
        (Xk.west)--(NW)|-(Xkdag.west)
        (Xkdag.east)--(A.west)
        (A.east)--(Xk1dag.west)
        (Xk1dag.east)-|(rhoS.south)
        (rhoS.north)--(rhoN.south)
        (rhoN.north)|-(Xk1.east)
        (Xk1.west)--(A2.east)
        (A2.west)--(Xk.east)
    ;
    \draw(A2.south)--(A.north);
\end{tikzpicture}

%% file: img/XArhovec.tex
\begin{tikzpicture}
    \node at(0,0)[draw,rectangle](A){$A$};
    \node at(-1.25,0)[draw,rectangle](Xk1dag){$X^{(k+1)\dag}$};
    \node at(1.,0)[draw,rectangle](rho){$\sqrt{\rho}$};
    \draw(A.north)--++(0,.5);
    \draw[ultra thick]
        (A.west)--(Xk1dag.east)
        (Xk1dag.west)--++(-.2,0)
        (A.east)--(rho.west)
        (rho.east)--++(.2,0)
    ;
\end{tikzpicture}

%% file: img/XAXrhovec.tex
\begin{tikzpicture}
    \node at(0,0)[draw,rectangle](A){$A$};
    \node at(-1.25,0)[draw,rectangle](Xkdag){$X^{(k)\dag}$};
    \node at(1.25,0)[draw,rectangle](Xk1dag){$X^{(k+1)\dag}$};
    \node at(2.5,0)[draw,rectangle](rho){$\sqrt{\rho}$};
    \draw(A.north)--++(0,.5);
    \draw[ultra thick]
        (A.west)--(Xkdag.east)
        (Xkdag.west)--++(-.2,0)
        (A.east)--(Xk1dag.west)
        (Xk1dag.east)--(rho.west)
        (rho.east)--++(.2,0)
    ;
\end{tikzpicture}

%% file: img/Xk1AAXk1vec.tex
\begin{tikzpicture}
    \node at(0,0)[draw,rectangle](AS){$A$};
    \node at(0,1)[draw,rectangle](AN){$A$};
    \node at(-1.25,0)[draw,rectangle](Xkdag){$X^{(k+1)\dag}$};
    \node at(-1.25,1)[draw,rectangle](Xk){$X^{(k+1)}$};
    \coordinate (NW) at ($(Xk.west)+(-.25,0)$);
    \draw[ultra thick]
        (AS.east)--++(.2,0)
        (AS.west)--(Xkdag.east)
        (Xkdag.west)-|(NW)--(Xk.west)
        (Xk.east)--(AN.west)
        (AN.east)--++(.2,0)
    ;
    \draw(AS.north)--(AN.south);
\end{tikzpicture}

%% file: img/Xk1AXk1AXkvec.tex
\begin{tikzpicture}
    \node at(0,0)[draw,rectangle](AS){$A$};
    \node at(0,1)[draw,rectangle](AN){$A$};
    \node at(-1.25,0)[draw,rectangle](Xkdag){$X^{(k)\dag}$};
    \node at(1.25,0)[draw,rectangle](Xk1dag){$X^{(k+1)\dag}$};
    \node at(-2.5,.5)[draw,rectangle](Xk1){$X^{(k+1)}$};
    \draw[ultra thick]
        (Xk1dag.east)--++(.2,0)
        (AS.east)--(Xk1dag.west)
        (AS.west)--(Xkdag.east)
        (Xkdag.west)-|(Xk1.south)
        (Xk1.north)|-(AN.west)
        (AN.east)--++(2,0)
    ;
    \draw(AS.north)--(AN.south);
\end{tikzpicture}

%% file: img/XXvec.tex
\begin{tikzpicture}
    \node at(0,0)[draw,rectangle](Xk1dag){$X^{(k+1)\dag}$};
    \node at(0,1)[draw,rectangle](Xk1){$X^{(k+1)}$};
    \coordinate (NW) at ($(Xk1.west)+(-.25,0)$);
    \draw[ultra thick]
        (Xk1dag.east)--++(.5,0)
        (Xk1dag.west)-|(NW)--(Xk1.west)
        (Xk1.east)--++(.5,0)
    ;
\end{tikzpicture}